\documentclass[showpacs,amsmath,showkeys,pra,twocolumn]{revtex4-1}
\usepackage{dcolumn}    
\usepackage{bm}         
\usepackage{ifpdf}
\usepackage{amssymb,lineno,amsfonts}
\usepackage{graphicx}   
\usepackage{bbm}
\usepackage{mathrsfs}
\usepackage{upgreek}
\usepackage{mathtools}
\usepackage{epstopdf}
\usepackage{setspace}
\usepackage{hyperref}
\usepackage[matrix,frame,arrow]{xy}
\usepackage{rotating}
\usepackage{float}
\usepackage{natbib}
\usepackage{color} 
\definecolor{med-blue}{RGB}{25,25,112} 
\hypersetup{colorlinks, linkcolor={red},citecolor={blue}, urlcolor={blue}}

\begin{document}
\title{Benford distributions in NMR}
\author{Gaurav Bhole, Abhishek Shukla, and T. S. Mahesh}
\email{mahesh.ts@iiserpune.ac.in}
\affiliation{Department of Physics and NMR Research Center,\\
Indian Institute of Science Education and Research, Pune 411008, India}

\begin{abstract}
{
Benford's Law is an empirical law which predicts the frequency of significant digits in databases corresponding to various phenomena, natural or artificial.  Although counter intuitive at the first sight, it predicts
a higher occurrence of digit 1, and decreasing occurrences to other larger digits.
Here we report the Benford analysis of various NMR databases and draw several interesting inferences.
We observe that, in general, NMR signals follow Benford distribution in time-domain as well as in frequency domain.
Our survey included NMR signals of various nuclear species in a wide variety of molecules in different phases, namely
liquid, liquid-crystalline, and solid.  We also studied the dependence of Benford distribution on NMR
parameters such as signal to noise ratio, number of scans, pulse angles, and apodization.
In this process we also find that, under certain circumstances, the Benford analysis can distinguish a genuine spectrum from a visually identical simulated spectrum. 
Further we find that chemical-shift databases and amplitudes of certain radio frequency pulses generated
using optimal control techniques also satisfy Benford's law to a good extent.
}
\end{abstract}
\maketitle

\section{Introduction}

Simon Newcomb, an American astronomer,  observed in 1881 that the first few pages of the logarithmic books wore off more than the last ones \cite{newcomb}. The initial pages of the book which contained logarithms of numbers beginning with the lowest digits were more referred to than the last ones. From these observations, Newcomb inferred that the digits 1 to 9 do not occur with equal probability in nature. However, Newcomb's article was not well recognized due to a lack of a mathematical structure. It was revisited by Frank Benford in 1938 with a mathematical formulation \cite{benford}.
He predicted that the probability $P_B(d)$ of first significant digit `d' in a given dataset to be
 \begin{equation}
P_B(d)=\log_{10} \left(1 + \frac{1}{d} \right).
\end{equation}
 This empirical law, known as `Benford's law', is a statistical inference, which predicts the non-uniform frequency distribution of most significant digits in a given set of data. Accordingly, frequency of the first significant digit `1' is as high as 30\%, whereas the larger digits occur with progressively decreasing frequencies, and the last digit `9' with a mere 5\% frequency. Of course, the unit sum of probabilities, i.e., $\sum_{d=1}^9 P_B(d) = 1$ can be verified easily.

The data satisfying Benford's Law can be obtained from a wide spectrum of sources varying from astrophysical \cite{Astrophysical1}, geographical \cite{geographical}, biological \cite{neural,biology2,aerobiological}, seismographic \cite{seismic}, and financial topics \cite{stockmarket,financial,financial1}. Violations of this law find bizarre applications such as detecting cases of tax fraud \cite{taxfruad} and election fraud \cite{electionfraud}.  Recently, Ujjwal and coworkers have
shown that Benford analysis is an efficient tool to study quantum phase transitions \cite{ujjwal1,ujjwal2}.

The existence of this empirical law has been attributed to the fact that the natural numbers we use are mere `ratios'
whereas the quantities that occur in nature are in exponents. 
The linear distribution of quantifying `digits' is only a human construct.  On the contrary, many of the natural processes are instead based on geometric series.  In other words, nature intrinsically `counts' in exponents. 
As a result, the Benford's law is scale as well as base invariant \cite{benford}. That means the distribution is unaffected even if the dataset is converted from, say, SI units to imperial units.  The distribution also holds in hexadecimal, octal, or any other number system as it does for decimal.

In this article, we have carried out Benford analysis of a number of databases related to NMR, mainly
the time and frequency domain NMR signals of a variety of nuclear species in various compounds existing
in liquid, liquid crystalline, and solid phases.
Further, we study the dependence of Benford distribution on NMR processing  parameters such as
apodization and acquisition parameters such as number of scans and pulse angles.
Our survey also includes chemical shifts and amplitudes of RF pulses.

We have organized this report in the following way.  The Benford analysis of various NMR signals is described
in section II.  In section III, we report the Benford analysis of chemical shift database and RF pulses.
Finally we conclude in section IV.

\section{Benford distribution in NMR signals}
First we shall describe the procedure that we adopted for the Benford
analysis.  The normalized real and imaginary parts of a phase-sensitive NMR signal,
in time domain or frequency domain, is concatenated to obtain a single column vector.
The signs of the elements in the vector are suppressed by taking absolute values.
Then we compute the distribution of all the
digits from 1 to 9 in the most significant place. The observed distribution
$P(d)$ is normalized such that $\sum_{d=1}^9 P(d) = 1$. To quantify
its distance from the expected distribution $P_B(d)$, we define a 
`Benford goodness parameter' (BGP) as
\begin{eqnarray}
\Delta P = \left( 1 - \sqrt{ \sum_{d=1}^{9} \frac{(P(d)-P_B(d))^2}{P_B(d)}} ~\right) \times 100.
\end{eqnarray}
For an ideal Benford distribution, $\Delta P = 100$, while for real-life distributions,
it can take lower values or even negative values.  In the following we survey
Benford distribution in databases related to NMR.

\begin{center}
\begin{figure}[t]
\includegraphics[trim=0cm 0cm 0cm 0cm, clip=true,width=8.8cm]{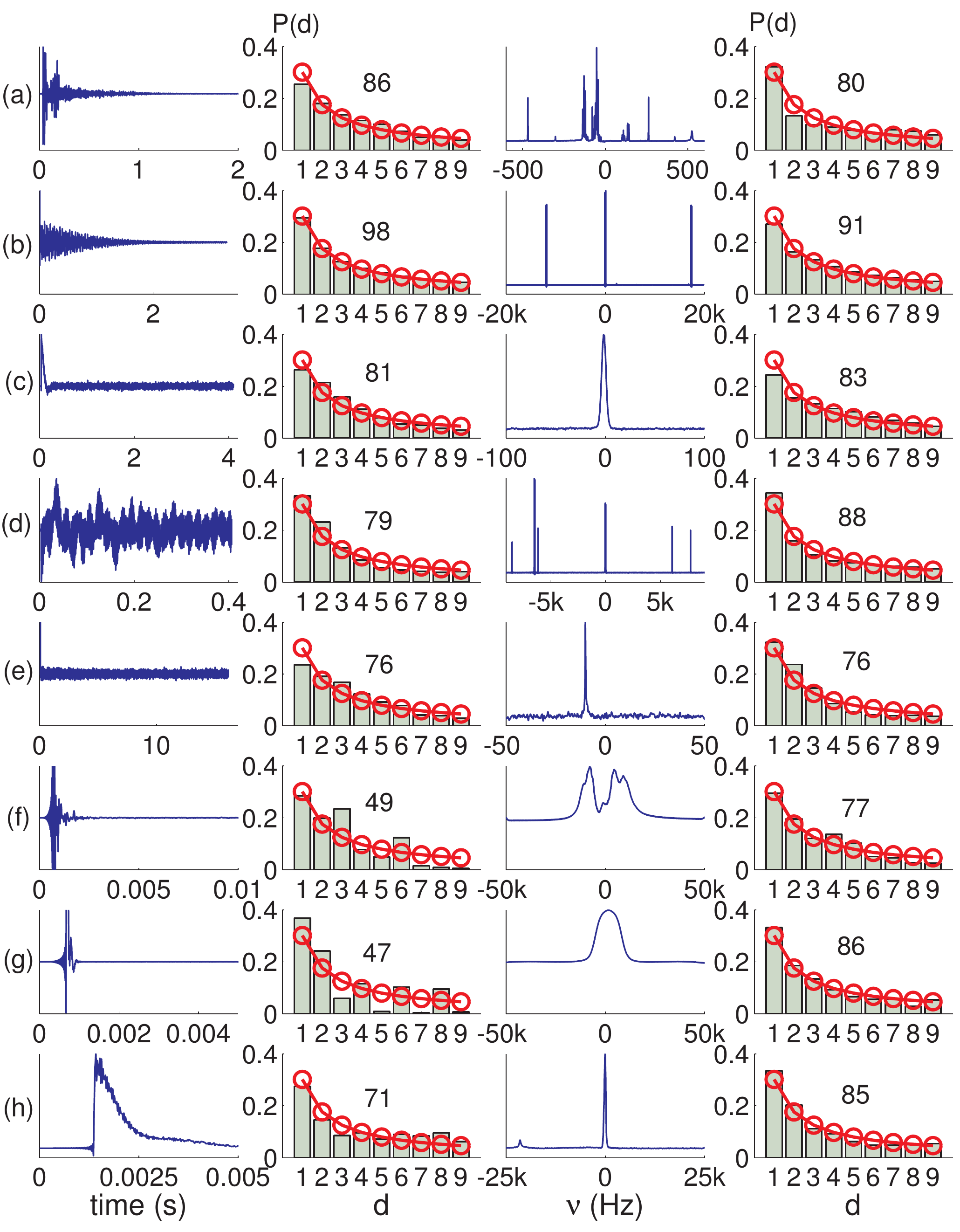} 
\caption{(Color online) Time-domain (1st column), and frequency-domain (3rd column) 
liquid state (rows a to e), liquid crystal (row f), and
solid state (rows g and h)  NMR signals, and their corresponding Benford distributions
(2nd and 4th columns) for
(row-wise)
(a) $^1$H of Diphenylphosphite,
(b) $^{19}$F of Trifluoroiodoethylene,
(c) $^{31}$P of Triphenylphosphine,
(d) $^{13}$C of Ethylbenzene, and
(e) $^{15}$N of Formamide.
(f) $^1$H of liquid crystal MBBA,
(g) $^1$H of Adamantane powder, and
(h) $^{13}$C Hexamethyl benzene.
The theoretical Benford distribution is indicated by circles connected by
lines. BGP value is also shown along with each bar plot.
}
\label{survey} 
\end{figure}
\end{center}

\subsection{A general survey}
Here we carry out Benford analyses of NMR signals obtained from various
nuclear species in different samples.
Figs. \ref{survey}(a-e) survey the Benford distribution of NMR signals from samples in liquid-state, while
Figs. \ref{survey}(f-h) survey those of certain samples in liquid crystalline and solid phases.
As evident from the high BGP values, we observe that NMR signals 
satisfy Benford's law to a great extent in both time and frequency domains.
The satisfaction in time domain is presumably due to the exponential
decay of the free induction signal, while the satisfaction in the frequency
domain is less intuitive.  
Interestingly, when compared to liquid samples, the BGP values in liquid crystalline and
solid samples are lower in time-domain, but comparable in frequency-domain.

In all the cases, irrespective of the physical state of the sample, the NMR signals
appear to satisfy Benford's law to a good extent.  Even when BGP value is less
than 50, the general trend shows the decreasing occurrence of ascending digits.
In all the cases that we studied, we found without exception that the digit `1'
accumulates the highest probability.  This indeed is a clear signature of emergence of 
Benford's distribution.

\begin{center}
\begin{figure}[b]
\includegraphics[trim=1cm 0cm 0cm 1cm, clip=true,width=7cm]{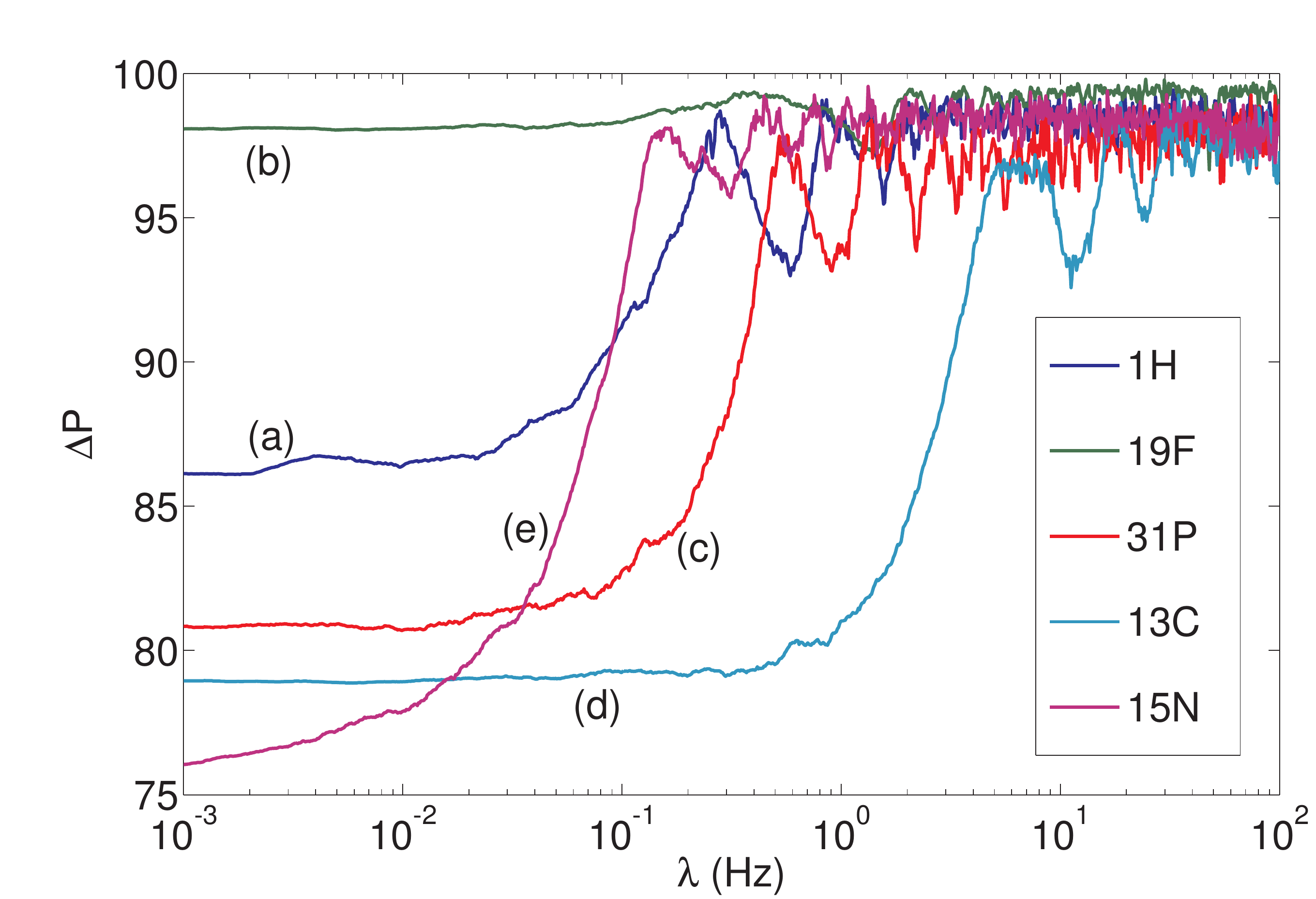} 
\caption{(Color online) BGP values versus the line width parameter $\lambda$
corresponding to the following liquid state NMR signals in time domain:
(a) $^1$H of Diphenylphosphite,
(b) $^{19}$F of Trifluoroiodoethylene,
(c) $^{31}$P of Triphenylphosphine,
(d) $^{13}$C of Ethylbenzene, and
(e) $^{15}$N of Formamide.
}
\label{apo} 
\end{figure}
\end{center}

\subsection{Effect of apodization}
In the following we study general behavior of BGP values under some of the 
acquisition or processing parameters.  First we describe the effect of 
apodization, i.e., multiplying the exponential window-function $e^{-\lambda t}$ 
with the time-domain dataset.  Here, $\lambda$ is the `line-width' parameter.
Such an apodization is routinely used in NMR processing for suppressing noise.
Fig. \ref{apo} shows the variation of BGP under apodization with increasing $\lambda$ values
for the liquid samples mentioned in Fig. \ref{survey}(a-e). 
Apodization appears to improve BGP value in all the cases, except 
in the case of $^{19}$F, where BGP already has a high value of $98$.
In each of the other cases, there appears to be a transition to a higher BGP value 
at a certain characteristic $\lambda$ value, and ultimately leading to the saturation
of BGP near 100.  These results clearly 
indicate that BGP improves with exponential apodization in general.

\subsection{Dependence on signal to noise ratio}
\begin{center}
\begin{figure}[b]
\includegraphics[trim=0cm 0cm 0cm 0cm, clip=true,width=6.5cm]{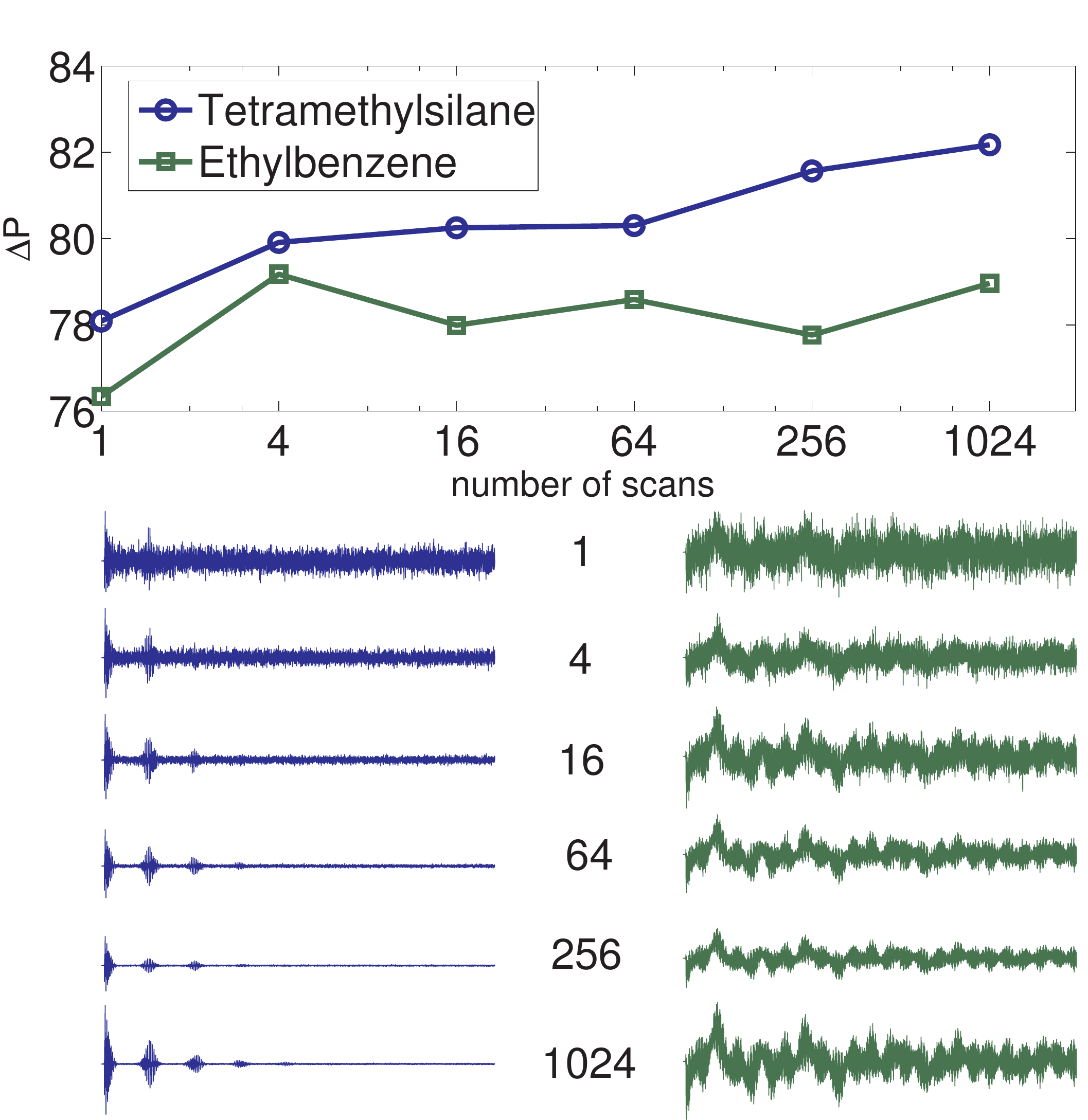} 
\caption{(Color online) BGP values versus the number of scans for
 time domain $^1$H NMR signals of liquid state samples (i) TMS and 
(ii) EB.  The real parts of the NMR signals of TMS (1st column) 
and EB (2nd column) are shown with arbitrary axes.
}
\label{sbyn} 
\end{figure}
\end{center}

To study the dependence of BGP on signal to noise ratio (SNR), we recorded 
different signals with varying number
of scans separately on two liquid-state samples (i) Tetramethylsilane (TMS)
and (ii) Ethylbenzene (EB).  
Since signal grows linearly with the number scans and noise grows as square root,
SNR also increases as square root of the number of scans \cite{cavanagh}. 
Therefore, measuring BGP values as a function of number of scans allows us to
study the SNR-dependence of BGP. The
results of the experiments are shown in Fig. \ref{sbyn}.  In the case of TMS, there 
appears to be a monotonic increase of BGP value with the number of scans, while
less significant improvement is seen in the case of EB.  These experiments
suggest the weak dependence of BGP on SNR.

\begin{center}
\begin{figure}
\includegraphics[trim=0cm 0cm 0cm 0cm, clip=true,width=7cm]{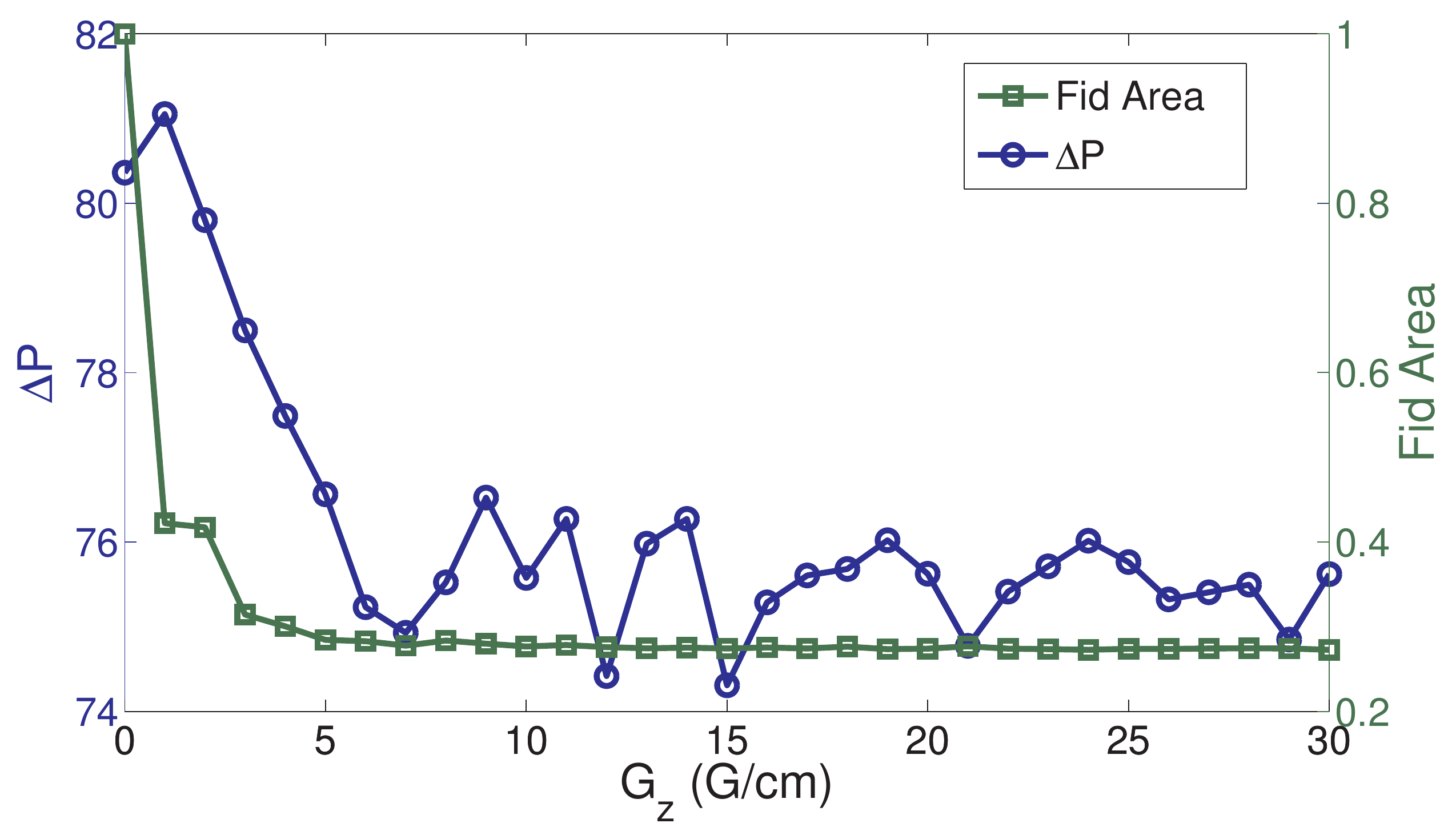} 
\caption{(Color online) BGP values (left axis) and FID area (right axis) versus
gradient strength $G_z$.
}
\label{pfg} 
\end{figure}
\end{center}

It is also possible to systematically decrease SNR
by the application of pulsed-field-gradients (PFGs), and monitor the effect on
the BGP values.  The results of the experiments on the liquid-samples of  TMS and EB 
are shown in Fig. \ref{pfg}.  Each of these points was obtained from the time-domain
$^1$H NMR signal after a 90 degree RF pulse followed by a PFG of 
strength varying between 0 to 30 G/cm.  With an increase in the PFG strength, one
observes a steady decrease in the overall transverse magnetization and corresponding 
decrease in FID area, as clearly seen in Fig. \ref{pfg}.  The BGP values also appear
to decrease with increasing PFG strength.  This is consistent with the previous 
discussion about Fig. \ref{sbyn}, that BGP and signal to noise ratio are directly related.

\begin{center}
\begin{figure}[b]
\includegraphics[trim=2cm 0cm 0cm 0cm, clip=true,width=8.8cm]{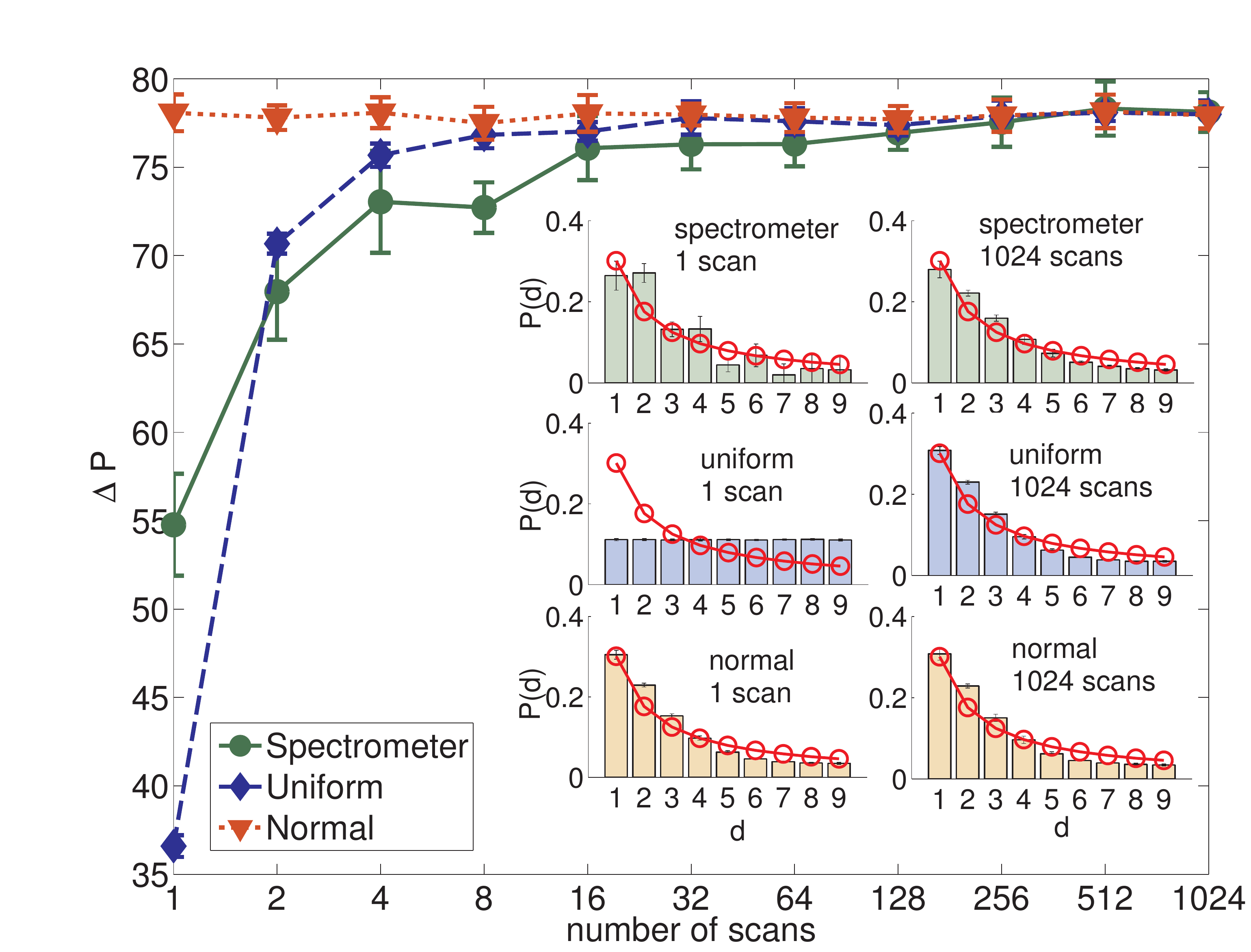} 
\caption{(Color online) BGP versus number of scans for spectrometer noise
(filled circles), simulated uniform noise (filled diamonds),
and simulated normally distributed noise (filled triangles).
The error bars are obtained using 10 independent datasets.
The bar plots displaying
Benford statistics of all three types of noises and for a single scan as well
as 1024 scans are shown in the inset.  
}
\label{noise} 
\end{figure}
\end{center}

\begin{figure*}
\includegraphics[trim=0cm 0cm 0cm 0cm, clip=true,width=18cm]{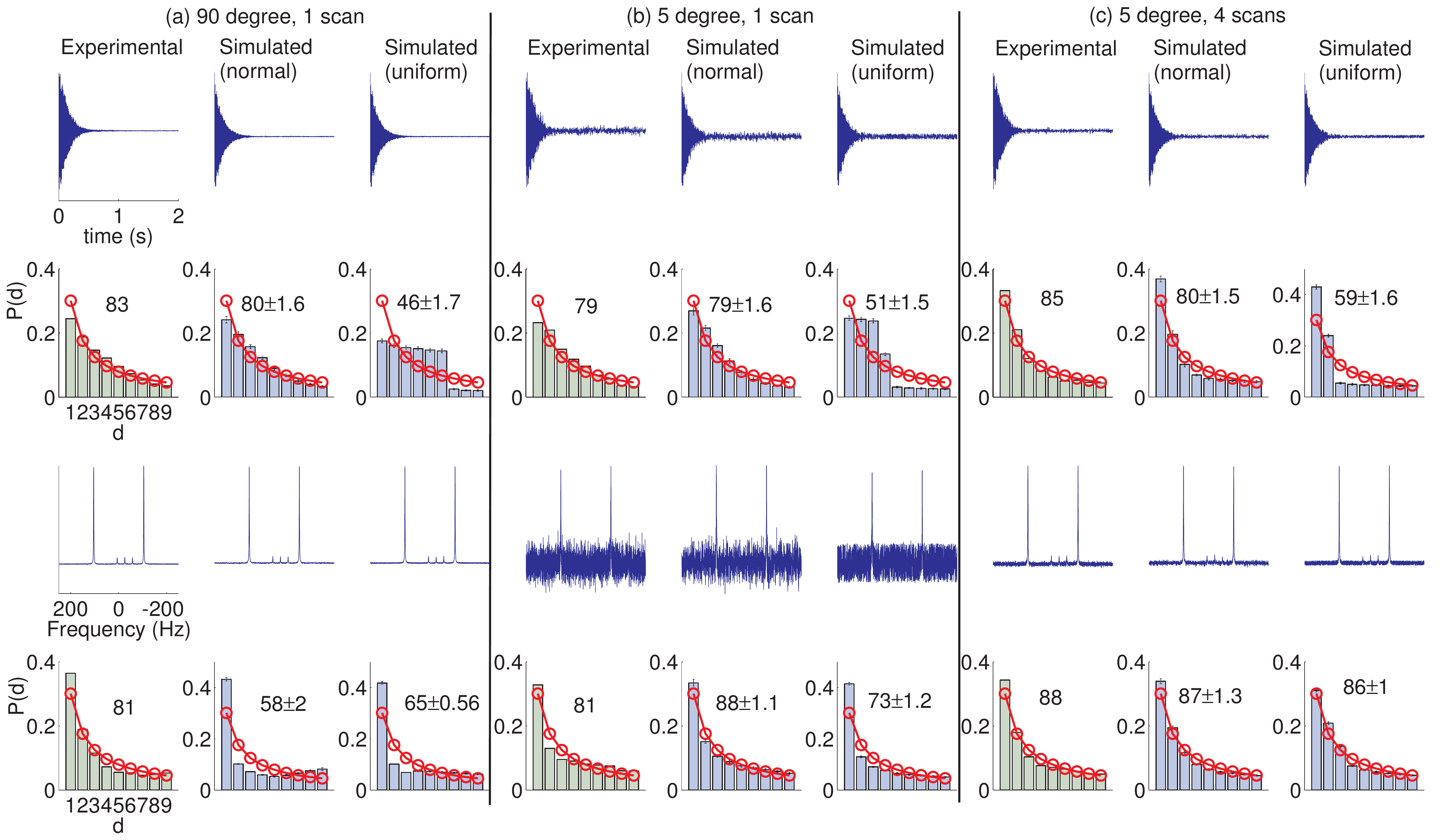} 
\caption{(Color online) 
NMR signals in time- (1st row) and frequency-domains (3rd row) of
experimental and simulated $^{13}$C NMR signals of $^{13}$CHCl$_3$
with varying pulse angles and number of scans.  Simulated signals
have appropriate levels of noises generated with `uniform'
as well as `normal' distributions as indicated.
Benford distributions
(2nd and 4th rows) are shown along with corresponding BGP values
under each signal. The theoretical Benford distribution is indicated
by circles with connecting lines.  The BGP error range of 
each simulated signal is obtained by a set of 10 independent datasets
with different noise vectors.
}
\label{chcl3}
\end{figure*}

\subsection{Noise averaging}
Now we shall describe the effect of averaging of noise on the BGP values. We 
recorded spectrometer noise (without sample) 
for a set of experiments with varying number of scans, and evaluated their
BGP values. The results are shown in Fig. \ref{noise}.  
The single scan spectrometer noise has a low BGP value of $55 \pm 3$, while
that of 1024 scans is 78 $\pm 1$. We observed that the BGP values
increase with noise averaging and saturate at $\Delta P \sim 80$.
We compare the experimental noise with two types of simulated noise: (i) 
uniformly distributed pseudorandom numbers and (ii) normally
distributed pseudorandom numbers.  Both types of pseudorandom numbers were
generated using Matlab (using functions $\tt{rand}$ and $\tt{randn}$ respectively), 
and normalized in the range $[0,1]$.
Noise averaging was then carried out by taking the mean vector of 
specific number (equal to number of scans) of random vectors.
The BGP values obtained for the mean vectors corresponding to various
number of scans are also shown in Fig. \ref{noise}.
Noise with uniform distribution display an equal probability for all the digits in
the most significant place as shown in the left-middle inset of Fig. \ref{noise}.  
This results in a very low BGP value of $36.57 \pm 0.05 $, which is close the 
theoretical value of $36.62$ for uniform occurrences of digits.
However averaging of uniform noise over multiple scans gives a characteristic distribution
close to Benford distribution (see right-middle inset). After 1024 scans, the BGP
value raised to $77.2$.
On the other hand, the normal distribution of noise has high BGP values ($77.9 \pm 0.3$) 
irrespective of the number of scans.  It is interesting to note that all the noises,
experimental as well as simulated, have similar BGP values (around 80) for large number
of scans.  

\subsection{Identifying genuine NMR signals}
With the above knowledge we can now try to distinguish a genuine (experimental)
NMR signal (in time- or frequency-domain) from a visually identical 
simulated NMR signal.  We choose the NMR signal from a simple model 
system, viz., $^{13}$C signal of $^{13}$CHCl$_3$ for our analysis.
We recorded its NMR signal with the following parameters 
(a) 90 degree pulse, 1 scan; (b) 5 degree pulse, 1 scan; 
and (c) 5 degree pulse, 4 scans.  We have also simulated the
corresponding time- and frequency-domain spectra using Matlab, 
by adding appropriate level of noise in each case so as to obtain
visually identical signals.  All the NMR signals and their Benford
distributions, along with their BGP values, are displayed in Fig. \ref{chcl3}.

By comparing all the data we draw the following observations.  All the 
experimental NMR signals had high BGP values (between 79 and 88).  
In the case (a), with high SNR, BGP values of all simulated signals 
are lower than that of experimental signals.  
Particularly, the simulated time-signal with uniform noise distribution
and frequency-signal with normal noise distribution had very low BGP values.
It is possible to clearly distinguish the experimental frequency-domain
signal form the simulated signals, although all the signals were visually similar.
In the case (b), with low SNR, the simulated signals with normal
distribution are indistinguishable from experimental signals in 
both time and frequency domains.  The simulated signals with uniform
distribution remain inferior with low BGP values.
In the case (c), with 4-scans signal averaging, 
all the signals are indistinguishable in frequency domain.
However, the BGP value of the simulated time-domain signal with uniform noise
has improved from the single scan case (as expected from Fig. \ref{noise}), 
but is still inferior compared to other signals. 

Thus we infer that Benford's law can be used in certain circumstances
to distinguish a genuine (experimental) NMR signal from a simulated one.

\section{Other Benford distributions in NMR}
So far we have considered Benford distributions of NMR signals. In the following
we study Benford distributions of chemical-shift database and of 
radio-frequency profiles.

\begin{center}
\begin{figure}
\includegraphics[trim=1cm 0cm 0.2cm 0cm, clip=true,width=7cm]{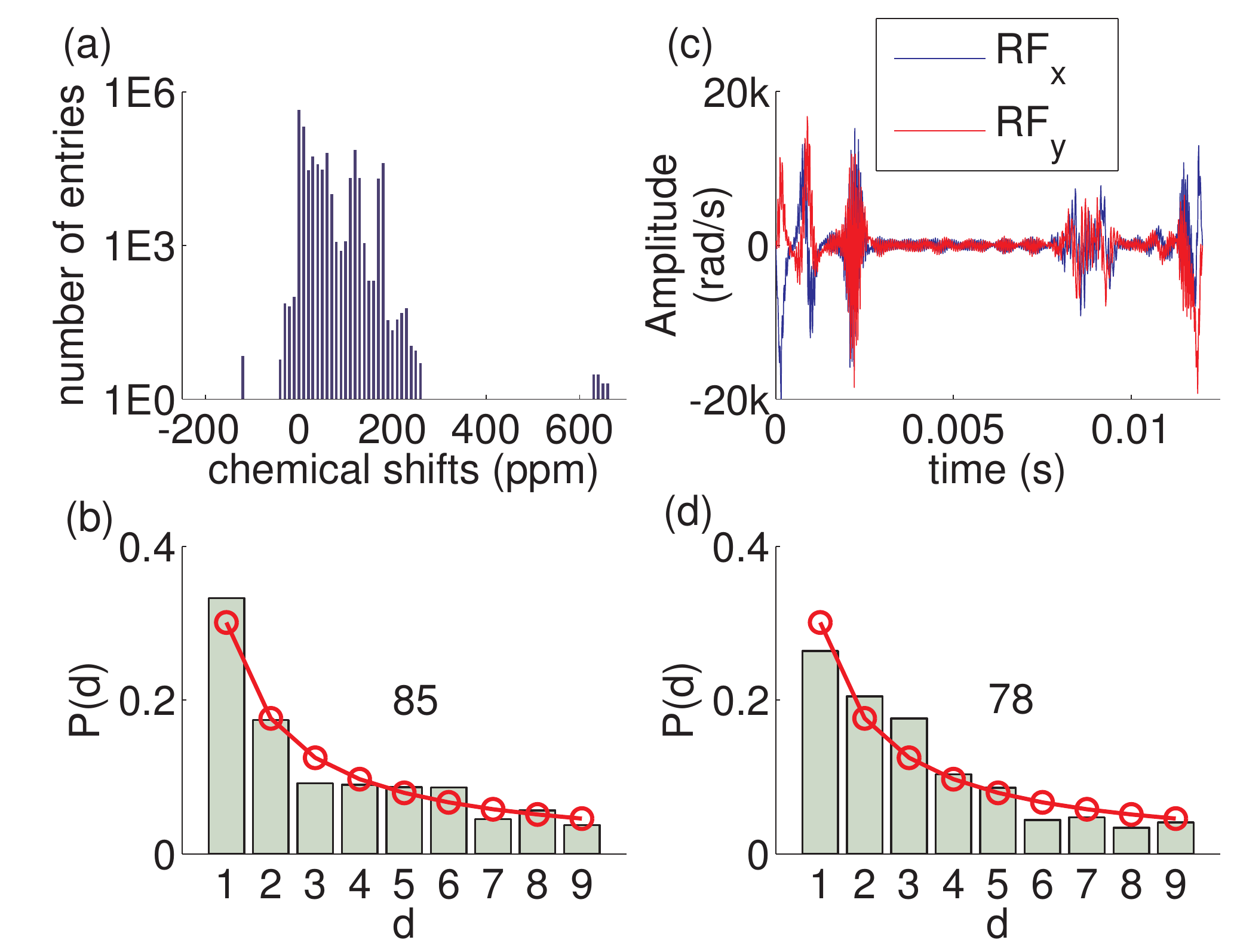} 
\caption{(Color online) (a) Number of entries in the BMRB chemical shift
database versus chemical shift.  Each bar corresponds to the number
of entries of various nuclear species in a range of 10 ppm.
(b) Benford distribution of the chemical shifts (normalized between 0 and 1).
The BGP value (85) is mentioned above the Benford distribution.
(c) The $X-$ (blue) and $Y-$ (red) components of RF amplitudes
of a GRAPE pulse corresponding to a CNOT gate.
(d) Benford distribution of the RF amplitudes normalized between 0 and 1.
The BGP value (78) is mentioned above the Benford distribution.
In (b) and (d), the theoretical Benford distribution is indicated by 
circles with connecting lines.
}
\label{csrf} 
\end{figure}
\end{center}

\subsection{Chemical shift database}
We downloaded a chemical shift database from Biological Magnetic Resonance data Bank (BMRB)
\cite{bmrb}.
The BMRB database is a repository for data from NMR spectroscopy on proteins, peptides, nucleic acids, and other biomolecules. The database contains over a million chemical shift entries of various nuclear species in
diverse biomolecules.  
The histogram of a number of entries
in every 10 ppm range from $-200$ ppm to $700$ ppm is shown in Fig. \ref{csrf}a.
Fig. \ref{csrf}b displays the Benford distribution for the chemical shift values
normalized between 0 and 1.  It is interesting to note the high BGP value
of 85 for this distribution.  

\subsection{Radio-frequency profiles}
Lastly, we report the Benford analysis of RF profiles which are generated from
optimal control techniques.  We used a popular numerical method, 
viz. Gradient Ascent Pulse Engineering (GRAPE) \cite{khaneja},
for generating high fidelity RF profiles. For example, a GRAPE RF profile for a CNOT gate in a three spin system
($^{19}$F of Trifluoroiodoethylene) having 800 segments, each of
duration 15 $\upmu$s and fidelity of over $0.97$, is displayed
in Fig. \ref{csrf}c.  We concatenate the $x-$ and $y-$ components of the GRAPE RF profile 
and normalize the absolute values between 0 and 1. The Benford distribution of 
the profile, along with its BGP value, is shown in Fig. \ref{csrf}d.  
Following table lists BGP values of some other GRAPE pulses generated
for different quantum operations. 
Reasonably high BGP values in most of these cases reveal the universality of Benford's law.\\

\begin{tabular}{|c|c|c|c|c|}
\hline
Pulse No. & Spin system & Segments & Fidelity & $\Delta P$ \\
\hline
\hline
1 & 3H,2F & 1500 & 0.98 & 86.4 \\
2 & 3H,2F & 1500 & 0.99 & 85.8 \\
3 & 3H,2F & 500 & 0.91 & 77.2 \\
4 & 3F & 500 & 0.95 & 85.7 \\
5 & 3F & 400 & 0.93 & 85.2 \\
\hline
\end{tabular}

\section{Conclusions}
Statistical analyses of datasets play an important role in many of the 
scientific studies.  Benford's analysis of a numerical dataset involves 
evaluating the distribution of the digits 1 to 9 in the most significant place,
and comparing the distribution with that predicted by Benford's law.  

In the present work we carried out Benford's analysis of NMR related databases.
These include time and frequency domains of NMR signals, a chemical-shift database,
and amplitudes of radio-frequency profiles generated by optimal control techniques.
We studied NMR signals from various nuclear species in a set of diverse molecules of 
samples in liquid, liquid-crystalline, and solid phases.  In all the cases, we 
observed a good agreement with Benford's law, and the digit `1' had
the most frequent occurrence without any exception.  We also studied the dependence
of Benford distribution under various NMR parameters such as signal to noise ratio, 
number of scans, pulse angles, and apodization.  We observed a weak dependence of the Benford
distribution on the signal to noise ratio.  

We also studied the Benford distributions of pseudorandom numbers with
uniform or normal distributions, and the effect of multiscan averaging.  
We also observed that under certain circumstances it is possible to
identify, using Benford analysis, a genuine NMR signal from a collection of 
visually identical simulated signals.
We believe that the present work is a first step in understanding the Benford
distribution in NMR related databases.

\section*{Acknowledgements}
The authors acknowledge Ujjwal Sen and Aditi Sen, of HRI, India, for bringing
Benford's law to their attention, and Anil Kumar of IISc, Bangalore, for
suggesting to carry out Benford analysis of NMR databases.
This work was partly supported by DST project SR/S2/LOP-0017/2009.

\bibliographystyle{apsrev4-1}
\bibliography{ref}
\end{document}